\documentclass[epj,nopacs,referee]{svjour}

\def\journal#1, #2, 1#3#4#5, #6#7    {
    {\rm #1~}{\bf #2}, #6#7 (1#3#4#5)}

\def\prl{\journal Phys. Rev. Lett., }

\def\npb{\journal Nucl. Phys. B, }
\def\plb{\journal Phys. Lett. B, }

\def\jmp{\journal J. Math. Phys., }
\def\jpa{\journal J. Phys. A, }
\def\zpc{\journal Z. Phys. C, }
\def\epjc{\journal Eur. Phys. J. C, }
\def\ajp{\journal Am. J. Phys., }

\def\Rb{{I\!\! R}}

\newcommand{\beq}[1]{\begin{equation}\label{#1}}
\newcommand\eeq{\end{equation}}
\newcommand{\ba}[1]{\begin{eqnarray}\label{#1}}
\newcommand{\baa}{\begin{eqnarray}}
\newcommand\ea{\end{eqnarray}}
\newcommand{\bee}{\begin{equation}}
\def\nn{\nonumber \\}
\def\l{\lambda}

\def\f{\varphi}

\newcommand{\cl}{Calogero}
\newcommand{\h}{Hamiltonian}

\def\hlf{\frac{1}{2}}
\begin{document}

\title{Deformed Heisenberg algebras, a Fock-space representation and the
Calogero model}

\author{Velimir Bardek\thanks{bardek@thphys.irb.hr} \and 
Stjepan Meljanac\thanks{meljanac@thphys.irb.hr} }

\institute{Theoretical Physics Division,
Rudjer Bo\v skovi\'c Institute, P.O. Box 180,
HR-10002 Zagreb, CROATIA}

\date{Received: date / Revised version: date}

\abstract{
We describe generally deformed Heisenberg algebras in one dimension. The 
condition  for a generalized Leibniz rule is obtained and solved. We analyze
conditions under which deformed quantum-mechanical problems have a
Fock-space representation. One solution of these conditions leads to a
q-deformed oscillator already studied by Lorek et al., and reduces to the 
harmonic oscillator only in the infinite-momentum frame. The other 
solution leads to the Calogero model in ordinary quantum mechanics, 
but reduces to the harmonic oscillator in the absence of deformation. }

\authorrunning{V. Bardek, S. Meljanac}
\titlerunning{Deformed Heisenberg algebras...}
\maketitle

\section{Introduction}

In the last few years considerable attention has been given to quantum groups 
acting on noncommutative spaces \cite{B1}. They represent a generalization 
of the concept of symmetries. A number of papers have established 
a connection between quantum groups, spaces, and q-deformed 
physics. The main idea behind these connections is that  the 
geometrical coordinates obey generalized (braid) statistics.
A specific approach to q-deformed phase space has been introduced by 
Schwenk and Wess \cite{W6} and Fichtm\"uller et al. \cite{W3}.
In the light of these ideas, the q-deformed harmonic oscillator has been 
studied \cite{W2}. Alternatively, the q-deformed harmonic oscillator based 
on an algebra of q-deformed creation and annihilation operators has been 
studied by Biedenharn and Macfarlane \cite{b,mcf}. 
A unified view of deformed single-mode oscillator algebras has been 
proposed \cite{bon,mmp}. 

A systematic approach to  q-deformed Heisenberg algebras
in one dimension and higher dimensions has been developed and applied to gauge 
field theories on noncommutative spaces \cite{w9,W1}. 
As the simplest example of  noncommutative structure, the one-dimensional 
q-deformed Heisenberg algebra has been considered \cite{W6}.
In addition, a differential calculus entirely based on the algebra has been 
derived. The laws of physics based on this calculus have been formulated. 
The representation of the algebra has been studied in the adapted 
quantum-mechanical scheme.

In this paper we generalize the one-dimensional Heisenberg algebra in such a 
way that the momentum in the $x$-representation is given by the function 
depending on one parameter or more parameters. General conditions on this 
function are imposed and, specially,
 the condition for a generalized Leibniz rule
is obtained and solved. Our main goal is to investigate under which 
condition the quantum-mechanical problem (deformed and undeformed) can be 
presented in the Fock-space representation. For q-deformed quantum mechanics 
there is only one solution constructed by Lorek et al. \cite{W2}, satisfying 
the Fock-space representation condition. However, we point out that this 
oscillator reduces to the harmonic one in the infinite-momentum frame. 

We have shown that there is only one solution satifying the Fock-space 
representation condition in generally deformed quantum mechanics (approaching 
the simple oscillator in ordinary mechanics). This solution can be 
interpreted as the Calogero model in the harmonic potential in 
ordinary quantum mechanics. Alternatively, the Calogero model can be 
viewed as deformed harmonic oscillator.

The paper is organized as follows. In Sec. 2 we describe generally 
deformed Heisenberg algebras in one dimension and discuss the generalized 
Leibniz rule. In Sec. 3 we analyze conditions under which deformed 
quantum mechanical problems have a Fock-space representation. 
In Sec. 4 we find that the Calogero model can be interpreted  as a 
deformed Heisenberg algebra with a Fock-space representation. 
Finally, in Sec. 5 we summarize the main result of the paper.

\section{Deformed Heisenberg algebras in one dimension}

In a  series of papers \cite{W6,W3,W2,W1,W4,W5,W7} a formal calculus entirely based on an 
algebra has been developed. Especially, the q-deformed Heisenberg algebra has 
been considered. The model is based on the q-deformed relations
\ba a
&& q^{\frac{1}{2}}xp- q^{-\frac{1}{2}}px=i\Lambda,\nn
&&\Lambda p=qp\Lambda, \nn
&& \Lambda x=q^{-1}x\Lambda, \;q\in\Rb,\;q\neq 0.\ea
Futhermore, the algebra  has to be a star algebra to allow a physical 
interpretation. The element $x$ of the algebra will
be identified with the observable for position in space, the 
element $p$ with the canonical conjugate observable (called momentum).
Observables have to be represented  by self-adjoint linear operators in 
 Hilbert space. This will guarantee real eigenvalues and a complete 
set of orthogonal eigenvectors.
Hence, the requirement  is
$$ \bar x=x,\;\bar p=p.$$ Then, for $q\in\Rb,\;q\neq 1$,
\ba b
px&=&i(q-q^{-1})^{-1}(q^{-\frac{1}{2}}\Lambda-q^{\frac{1}{2}}\bar\Lambda),\nn
xp&=&i(q-q^{-1})^{-1}(q^{\frac{1}{2}}\Lambda-q^{-\frac{1}{2}}\bar\Lambda).\ea
Futhermore, the algebra is extended by $\Lambda^{-1},\;x^{-1}$, and it is 
demanded that $$\bar\Lambda=\Lambda^{-1}.$$ A field $f$ is defined as an 
element of the subalgebra generated by $x$ and $x^{-1}$, and then
completed by a formal power series $f(x)\in{\cal A}_x$.
The above algebra ${\cal A}$, Eq.(\ref{a}), 
can be represented on ${\cal A}_x$ in a
natural way (the $x$-representation in deformed QM):
\ba c
&& x\rightarrow x,\;p\rightarrow -i\nabla, \nn
&& N=x\frac{d}{dx},\;\bar N=-\frac{d}{dx}x=-1-N,\;Nx^n=nx^n, \nn
&&\Lambda=q^{-N-\frac{1}{2}}.\ea
Then
\ba d
&&\bar\Lambda=q^{-\bar N-\frac{1}{2}}=q^{N+\frac{1}{2}}=\Lambda^{-1},\nn
&&x\nabla=\frac{q^N-q^{-N}}{q-q^{-1}}=\frac{\sinh(Nh)}{\sinh(h)},\;q=e^h,\nn
&&\nabla x=\frac{\sinh((N+1)h)}{\sinh(h)}.\ea
Let us generalize the above algebra and consider its representation in the 
following way:
\ba e
&&x\nabla=f_h(N),\nn
&&\nabla x=-{\overline{f_h(N)}}=-\bar f_h(\bar N)=-\bar f_h(-N-1)=f_h(N+1),\nn
&&{\rm if}\;\;\bar f_h(-N)=-f_h(N).\ea
Then $$\nabla=\frac{1}{x}f_h(N).$$
We consider $f_h(x)$ as a real function, with the restriction 
$f_0(N)=N$ as 
continuous deformation parameters $h\rightarrow 0$.
One easily finds
$$\nabla x^n=f(n)x^{n-1},\;n\in {\bf Z},\;f(1)=1,\;f(0)=0.$$
For example,
$$\nabla x =1,\;\nabla c =0,\;
\nabla f(x)=\nabla\sum_kc_kx^k=\sum_kc_kf(k)x^{k-1}.$$

Next, we consider what conditions the function $f$ should satisfy in order 
that a generalized derivative should satisfy the generalized Leibniz 
rule \cite{W1}:
\ba i
&&\nabla(x^{n+m})=(\nabla x^n)(\f(m)x^m)+(\f(-n)x^n)(\nabla x^m),\nn
\Longrightarrow\;&&f(n+m)=f(n)\f(m)+\f(-n)f(m).\ea
The functions $f(t),\;\f(t)$ should satisfy the following  conditions in 
the undeformed case:
\ba j
&&\lim_{h\to 0}\f_h(t)=1,\;\forall t,\nn
&&\lim_{h\to 0}f_h(t)=t.\nonumber\ea
These conditions suggest a simple ansatz $\f_h(t)=cf'_h(t)$, i. e.,
\ba k
&&\f_h(-t)=\f_h(t),\nn
&&f(t+u)=cf(t)f'(u)+cf'(t)f(u),\nn
&&f(2t)=2cf(t)f'(t),\;cf'(0)=1.\ea
The general solution is given by two one-parameter families:
$${\rm i)}\;\;f_h(t)=\frac{\sinh(th)}{\sinh(h)},\;\f_h(0)=\frac{ch}{\sinh(h)}=1,
$$considered by Wess et al. \cite{W3,W2,W1} (connected with the quantum-group 
consideration for $q\in{\Rb},\;q=\exp(h)$), and
$${\rm ii)}\;\;f_h(t)=\frac{\sin(th)}{\sin(h)},\;\f_h(t)=\cos(th). $$
To our knowledge, this latter family has not been considered and might 
be connected with the quantum group 
for $q\in{\bf S_1},\;q=\exp(ih),\;h\in\Rb$.\\
\underline{Remark 1}:\\
There are other solutions not covered by the ansatz (\ref{k}). 
For example,
\[ \f(t)=\left\{\begin{array}{c}
q^t\\q^{-t}\\\frac{q^t+q^{-t}}{2}=\cosh(th)\end{array}\;\right. {\rm for}\;
f_h(t)=\frac{\sinh(th)}{\sinh(h)} \]
and 
\[ \f(t)=\left\{\begin{array}{c}
q^{it}\\q^{-it}\\\frac{q^{it}+q^{-it}}{2}=\cos(th)\end{array}\;\right. 
{\rm for}\;
f_h(t)=\frac{\sin(th)}{\sin( h)}. \]
Other solutions of the initial Leibniz condition are not known to us at 
present.\\
\underline{Remark 2}:\\
For $q\in S_1,\; q=e^{ih},\;h\in\Rb$, the corresponding algebra is 
\baa
&& q^{\frac{1}{2}}xp- q^{-\frac{1}{2}}px=i\Lambda,\nn
&&\Lambda p=qp\Lambda ,\;\Lambda x=q^{-1}x\Lambda,\nn
&&\bar x=x,\;\bar p=p,\;\bar\Lambda=\Lambda=q^{-N-\frac{1}{2}},\nn
&&\nabla=\frac{1}{x}\frac{\sin(Nh)}{\sin(h)},\;\sin(h)\neq 0. 
\nonumber\ea
The general function $f(t)$, with the conditions $\bar f=f$, $f(t)-$odd, can be 
written as
\beq l
f(N)=\sum_{k=0}^{\infty}c_kN^{2k+1}.\eeq
Now, we define the \h \ in one dimension as
\baa
&&H=\hlf p^2+V(x)=-\hlf\nabla^2+V(x),\nn
&&H\Psi=i\frac{\partial}{\partial t}\Psi=E\Psi.\nonumber\ea
The simplest examples of dynamics are given by the following cases:\\
a)The free \h \ with $V(x)=0$,
\baa
&&\Psi(x)\sim\left[e^{ipx}\right]\equiv\sum_{n=0}^{\infty}
\frac{(ipx)^n}{[f(n)]!},\;
[f(n)]!=f(n)\cdots f(2)f(1),\nn
&&\nabla\left[e^{ipx}\right]_f=ip\left[e^{ipx}\right]_f,
\;p\in\Rb .\ea
b) The quadratic potential $V(x)=x^2$.\\
\underline{Remark 3}:\\
Orthogonality, completeness, and q-deformed integrals can be considered 
analogously as in Ref.\cite{W1}.

\section{Fock-space representation and one-dimensional quantum-mechanical
problems}

We consider a class of deformed quantum-mechanical problems in one 
dimension which allow the  Fock-space representation of a  deformed single 
(one-dimensional) oscillator. The \h \ is given by
\ba m
&&H=\hlf p^2+V(x)=-\hlf\nabla^2+V(x),\nn
&&\nabla=\frac{1}{x}f(N),\;N=x\frac{d}{dx} .\ea
We assume that the spectrum is discrete and bounded, i. e., that there 
exists a ground state with the lowest energy $E_0$:
\[ H\Psi_n=E_n\Psi_n,\; n\in N_0 . \]
Then we pose the question which class of potentials $V(x)$ (for the 
fixed deformation $f(N)$) leads to the deformed oscillator with the 
Fock-space representation in the following way:
\ba n
&&H=a^{\dagger}a+E_0,\nn &&a^{\dagger}a=E(N)-E_0,\nn && [N,a]=-a,\; 
[N,a^{\dagger}]=
a^{\dagger},\nn
&&a=F(p,x,N),\;\Psi_n=(a^{\dagger})^n|0\rangle,\;\Psi_0=|0\rangle,\nn
&&a|0\rangle=a\Psi_0=0, \ea
where $a^{\dagger}$ is the Hermitian conjugate of $a$,
$a^{\dagger}\neq a$.\\
\underline{Remark 4}:\\ 
There is a class of problems which can be mapped onto the oscillator problem
 \cite{BB,bor}. However, here we insist on the Fock-space representation.\\
If 
$$|n\rangle=(a^{\dagger})^n|0\rangle,\;\langle n|n\rangle >0,
\;\forall 
n\in N_0,\;a^{\dagger}a=E(N),$$
and 
$$ a^{\dagger}a(a^{\dagger})^n|0\rangle=E(n)(a^{\dagger})^n|0\rangle, $$
then
$$a(a^{\dagger})^n|0\rangle=E(n)(a^{\dagger})^{n-1}|0\rangle+|v\rangle,\;
a^{\dagger}|v\rangle=0.$$
This would imply that for some $p$
$$(a^{\dagger})^{p-1}|0\rangle\neq 0,\;{\rm and}\;(a^{\dagger})^{p}|0\rangle=0,
$$ in contradiction with our initial assumption that $|n\rangle\neq 
0,\;\forall n\in N_0$. Hence
$$a(a^{\dagger})^n|0\rangle=E(n)(a^{\dagger})^{n-1}|0\rangle ,$$
i. e., implying $$aa^{\dagger}=E(N+1).$$ Hence, the two Hermitian operators
$a^{\dagger}a$ and $aa^{\dagger}$ commute:
\bee\label{comm} [aa^{\dagger},a^{\dagger}a]=0.\eeq
The opposite is also true. If $aa^{\dagger}\neq a^{\dagger}a$, 
$[aa^{\dagger},a^{\dagger}a]=0$, and $a|0\rangle=0$, then we 
simultaneously  diagonalize the Hermitian operators $a^{\dagger}a$ and 
$aa^{\dagger}$. Assuming the discrete spectrum, we have
\ba p 
a^{\dagger}a|n\rangle&=&\f(n)|n\rangle,\nn
aa^{\dagger}|n\rangle&=&\psi(n)|n\rangle,\;n\in N_0,\ea
with $\f(n),\;\psi(n)$ bounded from below.
Starting from any state $|k\rangle $, we have
\baa 
a^{\dagger}a|k\rangle&=&\f_k|k\rangle,\nn
aa^{\dagger}|k\rangle&=&\psi_k|k\rangle.\ea 
Then$$a^{\dagger}a(a^{\dagger}|k\rangle)
=a^{\dagger}\psi_k|k\rangle=\f(k')a^{\dagger}
|k\rangle.$$ Since $[aa^{\dagger},a^{\dagger}a]=0$, then
$$aa^{\dagger}(a^{\dagger}|k\rangle)=\psi(k')a^{\dagger}|k\rangle.$$
In the same way we find
$$a^{\dagger}a(a^{\dagger}a^{\dagger}|k\rangle)=a^{\dagger}\psi(k')
a^{\dagger}|k\rangle=\f(k'')(a^{\dagger}a^{\dagger})|k\rangle, $$
and
$$aa^{\dagger}(a^{\dagger}a^{\dagger}|k\rangle)=\psi(k'')(a^{\dagger}
a^{\dagger})|k\rangle, $$
etc. Hence, the whole set of eigenstates can be split into a 
(finite or infinite)
set  of "bands" of the type
$\{(a^{\dagger})^n|k_1\rangle,...,\\(a^{\dagger})^n|k_{\alpha}\rangle,...
|n\in N_0\}$.\\
\underline{Remark 5}:\\
The procedure can go in the opposite direction:
\baa
aa^{\dagger}(a|k\rangle)&=&a\f(k)|k\rangle=\tilde\psi'_ka|k\rangle,\nn
a^{\dagger}a(a|k\rangle)&=&\tilde\f'_ka|k\rangle,\nn
aa^{\dagger}(aa|k\rangle)&=&a\tilde\f''(k)a|k\rangle=
\tilde\psi''(k)aa|k\rangle .\nonumber\ea
Hence, the necessary and sufficient condition for the one-dimensional 
quantum-mechanical problem to allow a Fock-space representation  is 
$[aa^{\dagger},a^{\dagger}a]=0,\;\langle 0|aa^{\dagger}|0\rangle>0,\;
(aa^{\dagger}\neq a^{\dagger}a$). If $aa^{\dagger}=a^{\dagger}a,\;a|0\rangle=0,$
there is no other state except $|0\rangle$.

\subsection{Undeformed (ordinary) quantum-mechanical problems in one dimension}

Let us apply the condition for the Fock-space representation (\ref{comm}) to 
undeformed quantum mechanics, with the following ansatz:
\baa \label{ab}
&&\sqrt 2a=ip+w(x)=\frac{d}{dx}+w(x),\nn
&&\sqrt 2a^{\dagger}=-ip+\bar w(x)=-\frac{d}{dx}+\bar w(x).\ea
This gives 
\baa\label{cd}
&&2a^{\dagger}a=-\frac{d^2}{dx^2}+|w(x)|^2-\frac{d}{dx}w(x)+\bar w(x)
\frac{d}{dx},\nn 
&&2aa^{\dagger}=-\frac{d^2}{dx^2}+|w(x)|^2+\frac{d}{dx}\bar w(x)-
w(x)\frac{d}{dx}.\ea
For the real function $w$, these equations lead to
\baa
&&\left[ H_0,\frac{dw}{dx}\right]=0,\;H_0=
-\hlf\frac{d^2}{dx^2}+\hlf w^2(x)^2=\hlf\{a,a^{\dagger}\},\nn
&&\frac{d^2}{dx^2}\left(\frac{dw}{dx}\right)-\frac{dw}{dx}\frac{d^2}{dx^2}=0,
\nn &&w'''+2w''\frac{d}{dx}=0,\nn
&&\frac{d^2w}{dx^2}=0\;\Longrightarrow\;w=\alpha x+\beta, \ea
where $\alpha,\;\beta\in\Rb$.\\
Next, we consider two special cases with the complex parameters $\alpha$ and 
$\beta$:
\baa
{\rm i)}\; \sqrt 2a&=&\frac{d}{dx}+\beta,\nn
\sqrt 2a^{\dagger}&=&-\frac{d}{dx}+\bar\beta,\nn
2a^{\dagger}a&=&-\frac{d^2}{dx^2}+|\beta|^2-\beta\frac{d}{dx}+\bar\beta
\frac{d}{dx},\nn H&=&\hlf\left(-\frac{d^2}{dx^2}+|\beta|^2\right)-i\Im\beta
\frac{d}{dx},\nn
\Re\beta&=&0\;\Longrightarrow\;aa^{\dagger}=a^{\dagger}a,\nn
\Im\beta&=&0\;\Longrightarrow\;aa^{\dagger}=a^{\dagger}a,\nn
2aa^{\dagger}&=&-\frac{d^2}{dx^2}+|\beta|^2+\bar\beta\frac{d}{dx}
-\beta\frac{d}{dx}=2a^{\dagger}a.\ea
Since $aa^{\dagger}=a^{\dagger}a$, and $a|0\rangle=0$, there is no Fock-space
representation.
\baa
{\rm ii)}\; \sqrt 2a&=&ip+\alpha x=\frac{d}{dx}+x\alpha,\nn
\sqrt 2a^{\dagger}&=&-ip+x\bar\alpha=-\frac{d}{dx}+x\bar\alpha,\nn
2a^{\dagger}a&=&-\frac{d^2}{dx^2}+|\alpha|^2x^2-\alpha\frac{d}{dx}x+x\bar\alpha
\frac{d}{dx},\nn
2aa^{\dagger}&=&-\frac{d^2}{dx^2}+|\alpha|^2x^2+\bar\alpha\frac{d}{dx}x
-x\alpha\frac{d}{dx},\nn
2a^{\dagger}a&=&2H_0-\alpha-2i\Im\alpha x\frac{d}{dx},\nn
2aa^{\dagger}&=&2H_0+\bar\alpha-2i\Im\alpha x\frac{d}{dx}=
2a^{\dagger}a+(\alpha+\bar\alpha).\ea
Hence, the Fock-space representation exists if $\alpha\neq 0$. In the 
limit $\alpha\to 0$, there is no Fock-space representation.

\subsection{q-deformed quantum-mechanical 
problems in one dimension}

Let us apply the conditions $[aa^{\dagger},a^{\dagger}a]=0,\;\langle 0|
aa^{\dagger}|0\rangle\neq 0$ to the operators $a,a^{\dagger}$ in (\ref{n}), 
which  satisfy  the q-deformed Heisenberg algebras, 
Eq. (\ref{a}) with $q\in\Rb$, (satisfying the Leibniz rule).
We generalize the above considerations with a natural (linear) 
restriction:
\bee\label{relq}aa^{\dagger}-u_qa^{\dagger}a=v_q,\;u_q,v_q\in\Rb,\eeq
and $u_q\rightarrow 1,\;v_q\rightarrow 1$ when $q\rightarrow 1$.\\
$[$Remark: Generally, it could be $aa^{\dagger}=\sum_{k=0}^{\infty}c_k
(a^{\dagger}a)^k$.$]$\\
Then the conditions $\langle 0|aa^{\dagger}|0\rangle>0$ and $[aa^{\dagger},
a^{\dagger}a]=0$ are obviously (automatically) satisfied if 
$v_q> 0$ (or $c_0>0$).\\
i) A simple ansatz is
\baa\label{az1} a&=&F_1(N)+F_2(N)p,\nn
a^{\dagger}&=&\bar F_1(-N-1)+p\bar F_2(-N-1).\ea
Then
\baa
aa^{\dagger}&=&F_1(N)\bar F_1(-N-1)+pF_2(N-1)\bar F_2(-N-2)p\nn
&+&pF_1(N-1)
\bar F_2(-N-1)+\bar F_1(-N-2)F_2(N)p,\nn
a^{\dagger}a&=&F_1(N)\bar F_1(-N-1)+pF_2(N)\bar F_2(-N-1)p\nn& +&pF_1(N)
\bar F_2(-N-1)+\bar F_1(-N-1)F_2(N)p,\ea
In order to obtain  relation (\ref{relq}), it is necessary that
\baa\label{nec}
F_1(N)\bar F_1(-N-1)&=&|\alpha|^2,\nn
F_2(N)\bar F_2(-N-1)&=&|\beta|^2,\;\alpha,\beta\in{\bf C}.\ea
A simple solution for  $F_2(N)$ is
\bee\label{sol1}
F_2(N)=\beta q^{N+\hlf},\;q>1 .\eeq
Then, in order to cancel the terms with $p^2$ and $p$, we obtain
\bee \label{sol1a} F_1(N)=\alpha q^{2(N+\hlf)},\eeq
since
\baa\label{sol1b}
&&F_2(N-1)\bar F_2(-N-2)=|\beta|^2q^{(N-\hlf)}q^{(-N-\frac{3}{2})}\nn
&&=q^{-2}F_2(N)\bar F_2(-N-1),\nn && F_1(N-1)=q^{-2}F_1(N).\ea
Hence the q-deformed algebra is identical with that of Refs.\cite{W2,b,mcf}:
\bee\label{nov}
aa^{\dagger}-q^{-2}a^{\dagger}a=\left(1-q^{-2}\right)|\alpha|^2=1,\;q>1.\eeq
Note that in the limit $q\rightarrow 1,\;\alpha=$const, we find
$aa^{\dagger}=a^{\dagger}a$ and there is no Fock-space representation for
$q=1$. 
The \h \ $H=a^{\dagger}a$ is given by
\baa\label{ham}
H&=&|\alpha|^2+|\beta|^2p^2+\alpha\bar\beta pq^{N+\hlf}+{\rm h.c.},\nn
H&=&\frac{1}{1-q^{-2}}+\hlf p^2+\frac{1}{\sqrt{2(1-q^{-2})}}pq^{N+\hlf}
+{\rm h.c.}. \ea
The vacuum state ($a|0\rangle_q=0$) is given by
\baa \label{vst}
&&\left(\alpha q^{2(N+\hlf)}+\beta q^{N+\hlf}p\right)\sum_{k=0}^{\infty}
c_kx^k=0,\nn
&&\alpha\sum_{k=0}^{\infty}c_kq^{2(k+\hlf)}x^k-i\beta\sum_{k=1}^{\infty}
c_kq^{k-\hlf}[k]x^{k-1}=0,\ea
where $$[k]=f(k)=\frac{q^k-q^{-k}}{q-q^{-1}}.$$
The recursion relations are
\baa\label{recur}
&&\alpha c_kq^{2(k+\hlf)}-i\beta c_{k+1}q^{k+\hlf}[k+1]=0,\nn
&&c_{k+1}=\frac{\alpha}{i\beta}\frac{q^{k+\hlf}}{[k+1]}c_k.\ea
We find that 
\bee\label{ck}
c_k=\left(\frac{\alpha}{i\beta}\right)^k\frac{q^{\frac{k}{2}}}{\{k\}!}c_0,
\;{\rm where}\;\{k\}=\frac{1-q^{-2k}}{1-q^{-2}}.\eeq
Hence the vacuum is
\baa \label{vcm}
&&|0\rangle_q=c_0\sum_{k=0}^{\infty}
\left(-\frac{i\alpha}{\beta}q^{\hlf}x\right)^k
\frac{1}{\{k\}!}=c_0\sum_{k=0}^{\infty}\frac{(ip_0q^{\hlf}x)^k}{\{k\}!},\nn
&&p_0=-\frac{\alpha}{\beta}.\ea
When $p$ is fixed and $q\rightarrow 1$, then 
$$|0\rangle_q\longrightarrow e^{ip_0x}.$$
However, in this limit we obtain
 the free \h \ boosted by $p_0=-\alpha/\beta$, and 
$aa^{\dagger}=a^{\dagger}a$, so there is no Fock-space representation.
Namely, in the limit $q\rightarrow 1,\;\alpha={\rm const}$, we obtain
$$(\alpha +\beta p)e^{ip_0x}=(\alpha +\beta p_0)e^{ip_0x}=0
\Longrightarrow p_0=-\frac{\alpha}{\beta},$$
and there is no Fock-space representation.

Only in the limit $|\alpha|=1/\sqrt{1-q^{-2}}\rightarrow\infty$, i. e., in the 
infinite-momentum frame, could one have the ordinary harmonic oscillator
picture \cite{W2}.
We show (see Appendix) that in the infinite-momentum frame:
\baa\label{re}
&&p_0=-\frac{\alpha}{\beta}=\pm\sqrt{\frac{2}{1-q^{-2}}},\nn
&&\lim_{p_0\to\infty}e^{-ip_0x}|0\rangle_q=c_0e^{-\frac{x^2}{2}},\nn
&&a=\pm\frac{1}{\sqrt{2}}x\pm i\frac{p}{\sqrt{2}},\;m=\omega=1.\ea
Generally, if $|\alpha|^2=1/(1-q^{-2}),\;q>1$, one obtains \cite{W2}
\baa\label{skor}
aa^{\dagger}&=&q^{-2(N_a+1)}{E_0}+\frac{1-q^{-2(N_a+1)}}{1-q^{-2}},\nn
a^{\dagger}a&=&q^{-2N_a}{E_0}+\frac{1-q^{-2N_a}}{1-q^{-2}},\ea
where 
$$[N_a,a]=-a,\;[N_a,a^{\dagger}]=a^{\dagger},$$
$$a^{\dagger}a|0\rangle=E_0|0\rangle=0,\;\lim_{n\to\infty}\langle n|
a^{\dagger}a|n\rangle=\frac{1}{1-q^{-2}}>0.$$
Hence, $E_0=0$ defines an ordinary  Fock-space representation 
with the bounded spectrum.
Any $E_0>1/(1-q^{-2})$ defines the representation with the unbounded 
spectrum  and with the  
states $(a^{\dagger})^n|E_0\rangle,\;a^n|E_0\rangle, \;n\in N_0$. 
If $E_0<1/(1-q^{-2})$, the eigenvalues of the  states $a^n|E_0\rangle$ 
would have 
values unbounded  from below for $n\rightarrow\infty$. Only if
$E_0=(1-q^{-2n})/(1-q^{-2}),\; n\in N_0$, then the lowest eigenvalue 
is zero. The limit $q=1$  reduces to the case of ordinary harmonic oscillator 
with $a^{\dagger}a=N$.\\
ii) In the q-deformed quantum mechanics \cite{W1,W2} defined by Eq.(\ref{d})
 we try to find such a Fock-space representation that leads to the 
ordinary harmonic oscillator picture in the limit $q\rightarrow 1$. Hence, 
we take a simple ansatz in the following form:
\baa \label{222}
a&=&f(N)x+ig(N)p,\nn
a^{\dagger}&=&x\bar f(-N-1)-ip\bar g(-N-1),\ea
where $f(N),\;g(N)$ are complex functions and for $q\rightarrow 1$ 
these functions tend to $1/\sqrt 2$. We find that 
\baa\label{333}
a^{\dagger}a&=&x\bar f(-N-1)f(N)x+p\bar g(-N-1)g(N)p\nn&+&ix\bar f(-N-1)g(N)p
-ip\bar g(-N-1)f(N)x,\nn
aa^{\dagger}&=&f(N)x^2\bar f(-N-1)+g(N)p^2\bar g(-N-1)\nn&+&ig(N)px\bar f(-N-1)
-if(N)xp\bar g(-N-1).\ea
A simple analysis shows that in any linear combination of 
$aa^{\dagger}$ and $a^{\dagger}a$ one cannot eliminate all terms on the 
right-hand  side of Eqs. (\ref{333}) in such a way as to obtain a 
constant. 
Namely, one can eliminate the terms with $x^2,\;p^2$, but the terms 
with $xp$ (or $px$) remain (unless $f,\;g$ are constants and $p=-i\frac{d}{dx}$
). Generally, one can simultaneously eliminate only three of the four 
terms $(x^2,p^2,xp,px)$,
and the condition $[aa^{\dagger},a^{\dagger}a]=0$ is not satisfied.
For example, if we take the \h \ with a quadratic potential, we have
\baa
&&p=-i\nabla=\frac{1}{x}\frac{\sinh(hN)}{\sinh(h)},\nn
&& \sqrt 2 a=x+ip,\; \sqrt 2 a^{\dagger}=x-ip,\nn
&& H=\hlf\{a,a^{\dagger}\}.\nonumber\ea
Since  $[aa^{\dagger},a^{\dagger}a]\neq 0$, there is no 
Fock-space representation.
Therefore, in the q-deformed quantum mechanics \cite{W1,W2} given by
Eq.(\ref{222}) there is no deformed  harmonic oscillator 
that goes to  $\sqrt 2 a=
x+ip$ in the limit $q=1$.
Having in mind the results of subsection B, it seems 
that there is  one q-deformed oscillator, Eq.(\ref{az1}),
that leads to the ordinary harmonic 
oscillator only in the infinite-momentum frame, Eq.(\ref{re}) 
(otherwise it leads to the 
free \h \ without a Fock-space 
representation). Our conclusion is that for the deformed momentum 
$p=-i\nabla$, with 
$\nabla$ satisfying the Leibniz rule (\ref{i}), 
there is no natural generalization of the  
creation and annihilation operators $a^{\dagger}$ and $a$ leading to the 
ordinary harmonic oscillator operators, Eq.(\ref{re}), in the limit 
$q\rightarrow 1$.

\section{Calogero model as a deformed Heisenberg algebra with a Fock-space
representation}

Let us now investigate the structure of the $\nabla$ operator  in 
order to get the deformed Heisenberg algebra with a Fock-space representation. 
 Assuming that the creation and annihilation operators depend linearly  on $x$
and $\nabla$ (as in the harmonic oscillator model),
\ba A
a&=&\frac{1}{\sqrt{2}}(\nabla +x),\nn
a^{\dagger}&=&\frac{1}{\sqrt{2}}(-\nabla +x), \ea
and insisting on the consistency condition (\ref{comm}),
we easily obtain
\beq C
[-\nabla^2+x^2,f(N+1)-f(N)]=0,\eeq
where the function $f$ has been introduced  in Eq.(\ref{e}). Consequently, the 
function $f$ 
is an identity
\beq D f(N)=N, \eeq
or satisfies the condition
\beq E f(N+3)-f(N+2)=f(N+1)-f(N) .\eeq
The first possibility  leads to the well-known case of  undeformed
harmonic oscillator, while the second one gives
\beq F f(N)=N+h\sin(\pi N) .\eeq
The $\nabla$ operator can now be expressed as 
\beq G \nabla=\partial +\frac{h}{x}\sin(\pi N) .\eeq
Note that the $\nabla$ operator does not satisfy the generalized 
Leibniz rule (\ref{i}).
The annihilation and creation operators can be rewritten as
\ba H
a&=&\frac{1}{\sqrt{2}}\left(\partial +\frac{h}{x}\sin(\pi N)+x\right),\nn
a^{\dagger}&=&\frac{1}{\sqrt{2}}\left(-\partial 
-\frac{h}{x}\sin(\pi N)+x\right). \ea
The \h \ can now be expressed as 
\ba I
H&=&a^{\dagger}a=
-\frac{1}{2}\partial^2+\frac{x^2}{2}-\frac{1}{2}+h
\sin(\pi N)+\frac{h}{2x^2}\sin(\pi N)\nn&+&\frac{h^2}{2x^2}\sin^2(\pi N).
\ea
This is in fact the \cl \ two-body \h \ \cite{C} with the  harmonic potential, 
where $x$ denotes the relative coordinate $x=x_1-x_2$. Namely, the action 
of the 
$\sin(\pi N)$ operator on the single power of $x$, say $x^{\l}$,
reduces to 
\beq J
\sin(\pi N)x^{\l}=\sin(\pi\l)x^{\l}. \eeq
It can be shown that the eigenfunctions of the \h \ are given by
\beq K
\psi_n(x)=x^{\nu+n}e^{-x^2/2}=(a^{\dagger})^n\psi_0(x)  \eeq
 and the corresponding eigenvalues by
\beq L \epsilon_n=n+\nu[1-(-)^n],\eeq
where $\nu=-h\sin(\pi\nu)$ and $n$ is an integer, $n\geq0$ \cite{pl}.
This gives for the function $f$:
\beq M 
f(N)=x\nabla=N+h\sin(\pi N)=N-\nu\frac{\sin(\pi N)}{\sin(\pi\nu)}.\eeq
By acting on the eigenfunctions (\ref{K}), we can  further 
simplify Eq.(\ref{M}) to
\beq N
x\nabla=N-\frac{\nu\sin(\pi(\nu+n))}{\sin(\pi\nu)}=N-\nu(-)^n=N-\nu K, \eeq
where $K$ denotes the so-called exchange operator, which, in our case,
 reduces to 
the sign flip. 

Let us now generalize the above results to the case of $N$ particles.
We shall postulate that the $N$-particle version of the deformed derivation 
(\ref{N}) is given by
\beq O x_{ij}\nabla_{ij}=N_{ij}-\nu K_{ij} \eeq
for each pair of indices, where
\ba P
 x_{ij}&=&x_i-x_j,\nn
N_{ij}&=&x_{ij}\frac{\partial}{\partial x_{ij}}.\ea
Here, $K_{ij}$ denotes the particle permutation operator, obeying
\ba X K_{ij}x_{ij}&=&-x_{ij}K_{ij},\nn K_{ij}&=&K_{ji},\nn
K_{ij}K_{jl}=K_{jl}K_{li}&=&K_{li}K_{ij},\nn K_{ij}^2&=&1. \ea
The operators $\nabla_{ij}$ satisfy the following commutation relations:
\baa\label{nova}
&&[\nabla_{ij},x_{ik}]=\delta_{jk}(1+\nu K_{ij})+\nu K_{ij},\nn
&&[\nabla_{ij}, \nabla_{ik}]=\frac{\nu}{x_{ij}}K_{ij}\left(\frac{\partial}
{\partial x_{jk}}-\frac{\partial}{\partial x_{ik}}\right)
\nn &&\hspace{1.8cm}+\frac{\nu}{x_{ik}}K_{ik}
\left(\frac{\partial}{\partial x_{ij}}-\frac{\partial}{\partial x_{kj}}\right)
\nn &&\hspace{1.8cm}+ \nu^2\left(\frac{1}{x_{ij}x_{jk}}
-\frac{1}{x_{ik}x_{kj}}\right)K_{ij}K_{ik}.\ea
Other commutators (i. e., commutators between the objects with 
pairwise noncoinciding indices) can be calculated from the 
set given above (\ref{nova}).
Furthermore, we define that the covariant derivative with respect to the
 i-th particle is given by
\beq R \nabla_i=\sum_{j\neq i}^N\nabla_{ij} .\eeq
The transition from the original set of variables $\{x_i\}$ to the new set 
of independent variables $\{x_{ij},X\}$, where $X$ denotes the center-of-mass 
coordinate,
\beq Q X=\frac{1}{N}\sum_i^Nx_i, \eeq
enables us to rewrite the sum (\ref{R}) as
\beq S \nabla_i=\sum_{j\neq i}^N\nabla_{ij}
=\frac{\partial}{\partial x_i}-\frac{1}{N}
\frac{\partial}{\partial X}-\nu\sum_{j\neq i}\frac{1}{ x_{ij}}K_{ij}. \eeq
In order to get rid of the center-of-mass degree of freedom, we 
finally redefine the covariant derivative (\ref{R}) as
\beq T D_i=\nabla_i+\frac{1}{N}\frac{\partial}{\partial X} .\eeq
Having in mind (\ref{nova}), it is easy to see 
that the  covariant derivation $D_i$ and the coordinate $x_i$ 
satisfy the commutation relations
\ba U &&[D_i, D_j]=0,\nn &&[x_i,x_j]=0, \nn 
&&[D_i,x_j]=\delta_{ij}\left(1+\nu\sum_{l=1}^NK_{il}\right)-\nu K_{ij},\ea
identical with those given in the paper \cite{Brink}.

The creation and anihilation operators can be constructed as
\ba V &&a^{\dagger}_i=\frac{1}{\sqrt{2}}(-D_i+x_i),\nn
&&a_i=\frac{1}{\sqrt{2}}(D_i+x_i) .\ea
The \h \ can  now be expressed as
\beq Z H=\sum_{i=1}^N a^{\dagger}_ia_i.\eeq
From a series of algebraic manipulations, using the properties of the particle 
permutation operator $K_{ij}$ (\ref{X}), the
\h \ (\ref{Z}) takes the form
\ba Y 
H&=&\frac{1}{2}\sum_{i=1}^N(-D_i^2+x_i^2)-\frac{1}{2}\sum_{i=1}^N[D_i,x_i]\nn
&=&-\frac{1}{2}\sum_{i=1}^N\partial_i^2+\frac{1}{2}\sum_{i=1}^Nx_i^2
+\frac{1}{2}\sum_{i\neq j}^N\frac{\nu(\nu-K_{ij})}{(x_i-x_j)^2}\nn&-&
\frac{\nu^2}{2}\sum_{i\neq j\neq l}^N\frac{1}{(x_i-x_j)(x_j-x_l)}K_{ij}K_{jl}
\nn&-&\frac{1}{2}\sum_{i=1}^N[D_i,x_i]. \ea
Finally, using the identity
\bee
\sum_{i\neq j\neq l}^N\frac{1}{(x_i-x_j)(x_j-x_l)}=0 \eeq
and the commutation relation (\ref{U}), we obtain the \h \ (\ref{Y}) in the form
\ba h
H=&-&\frac{1}{2}\sum_{i=1}^N\partial_i^2+\frac{1}{2}\sum_{i=1}^Nx_i^2
+\frac{1}{2}\sum_{i\neq j}^N\frac{\nu(\nu-K_{ij})}{(x_i-x_j)^2}\nn
&-&\frac{1}{2}\sum_{i=1}^N(1+\nu\sum_{j\neq i}^NK_{ij}).\ea
Noting that the operator $K_{ij}$ gives $1$ and $(-1)$ in the bosonic 
and fermionic subspace, respectively, we finally obtain 
\ba f
H=&-&\frac{1}{2}\sum_{i=1}^N\partial_i^2+\frac{1}{2}\sum_{i=1}^Nx_i^2+
\frac{1}{2}\sum_{i\neq j}^N\frac{\nu(\nu\mp 1)}{(x_i-x_j)^2}\nn
&-&\frac{N}{2}\mp\frac{\nu}{2}N(N-1), \ea
where the upper and lower sign refers to the symmetric and 
antisymmetric functions, respectively.
This is simply  the Calogero model for $N$ particles in the 
harmonic potential with the frequency $\omega$ equal to $1$ \cite{Poly92}.

\section{Conclusion}

We have generalized the  q-deformed one-dimensional Heisenberg 
algebra 
 \cite{W6,W3,W2,W1,W4,W5,W7}, to arbitrary deformations. We have  discussed 
 the conditions leading 
to a generalized Leibniz rule. We have  found two classes of solutions:
 one already studied 
by Wess et al. \cite{W6,W3,W2,W1,W4,W5}
 and the other 
 for $q=\exp(ih),\;h\in\Rb$.
Furthermore, we have given 
a simple condition that the one-dimensional problem should 
have a Fock-space 
representation. In ordinary Q.M., this condition leads to the 
ordinary harmonic oscillator. We have applied it to the special deformation
$\nabla=\sinh(hN)/x\sinh(h)$, $N=xd/dx$, considered in 
Refs.\cite{W6,W3,W2,W1,W4,W5,W7}. We
have found  only one solution leading to a q-deformed oscillator \cite{W2},
which reduces to the harmonic oscillator only in the infinite-momentum
frame.

Finally, we have imposed the condition for a Fock-space representation on 
$a\sim\nabla+x,\;a^{\dagger}\sim-\nabla+x$ and found that the solution for 
$\nabla$ is $\sim N+h\sin(\pi N),\;N=xd/dx$. This leads to the 
two-particle  Calogero model in ordinary Q.M. We have also found  the 
generalization of the $N$-particle Calogero model in ordinary Q.M.
It is interesting that we have found no other example of deformed 
quantum mechanics with a Fock-space representation.
It would be
interesting to find the corresponding two-dimensional oscillator in the
quantum plane.

{\bf Acknowledgement}

We would like to thank I. Dadi\'c, D. Svrtan and L. Jonke for useful
discussions.
This work was supported by the Ministry of Science and Technology of the
Republic of Croatia under Contract No. 00980103.

\appendix
\section{Appendix}

Expanding the plain wave in  power series and using the structure of
the vacuum (\ref{vcm}), we have
\bee\label{st1}
e^{-ipx}|0\rangle_q=c_0\sum_{n=0}^{\infty}(-i)^np^n_0x^n\sum_{k=0}^{n}
\frac{(-)^k}{(n-k)!\{k\}!}q^{\frac{k}{2}}.\eeq
To control the infinite momentum $p_0$, we expand it around the
undeformed point $q=1+h$, with $h$ as an infinitesimal small
quantity. The momentum $p_0$ goes to infinity as
$$p=-\frac{1}{\sqrt{h}}.$$
To proceed, we have to expand the sum
$$f_n(q)=\sum_{k=0}^{n}\frac{(-)^k}{(n-k)!\{k\}!}q^{\frac{k}{2}}$$
into a power series of $h=q-1$:
$$f_n(q)=\left.\sum_{m=0}^{\infty}\frac{h^m}{m!}f_n^{(m)}(q)
\right|_{q=1}.$$
It is easy to see that the $m-$th derivative of the function $f(q)$ with
respect to $q$ is given by
 $$f_n^{(m)}(1)=\sum_{k=0}^{n}\frac{(-)^k}{(n-k)!k!}P_{2m}(k), $$
where $P_{2m}(k)$ is some polynomial in $k$ of order $2m$. Using
the identity \cite{B2}
$$\sum_{k=0}^{n}\frac{(-)^k}{(n-k)!k!}S_m^{(k)}=0,\;{\rm for}\;n>m,$$
we find that $f_n^{(m)}(1)$ vanishes for $m<n/2$ and $n$ even,
and  it is different from zero for $m\geq n/2$.
For $n$ odd, say $n=2l+1$, $f_n^{(m)}(1)$ vanishes for $m\leq l$
and the first nonvanishing value appears for $m=l+1$.
Consequently, $e^{-ip_0x}|0\rangle_q$ is given by
\baa\label{ress}
e^{-ip_0x}|0\rangle_q&=&c_0\sum_{n=0}^{\infty}(i)^nh^{-\frac{n}{2}}x^n
\sum_{m=0}^{\infty}\frac{h^m}{m!}f_n^{(m)}(1)\nn&=&
c_0\sum_{n=0}^{\infty}(i)^nx^n\sum_{m=0}^{\infty}\frac{h^{m-\frac{n}{2}}}
{m!}
f_n^{(m)}(1).\ea
For $n$ odd, this expansion goes to zero as $\sqrt{h}$, while for
$n$ even, reduces to
\bee\label{ree}
c_0\sum_{n=0}^{\infty}(i)^nx^n\frac{f_n^{\left(\frac{n}{2}\right)}(1)}
{\left(\frac{n}{2}\right)!}.\eeq
Since $\{r\}_{q=1}^{(m)}$ goes like $r^{m+1}$, the dominant term of
$f_n^{(m)}$ is given by
\bee\label{dod}
\left(\sum_{r=1}^k\frac{\{r\}'}{\{r\}}\right)^m\frac{q^{k/2}}{\{k\}!},
\; m\leq k\leq n .\eeq
One easily finds that the sum over $r$ in (\ref{dod}) is given by
$$\sum_{r=1}^k\frac{\{r\}'}{\{r\}}=-\frac{k(k-1)}{2}.$$
Hence, it follows that
\bee\label{joss}f_n^{\left(\frac{n}{2}\right)}(1)=\frac{1}{2^{\frac{n}{2}}}.\eeq
Substitution of the expresssion (\ref{joss}) into  (\ref{ree}) 
results in the Taylor's expansion
for $\exp(-x^2/2)$.


\thebibliography{99}
\bibitem{B1}
C. Kassel, {\it Quantum Groups}, (Springer, Berlin 1995);
G. Lusztig, {\it Introduction to quantum groups}, (Progress in mathematical
physics, vol.110, Birkh\"auser 1993);
M. Chaichian, A. Desnichev, {\it Introduction to quantum groups},
(World Scientific, Singapore 1996).
\bibitem{W6}
A. Schwenk and J. Wess, \plb 291, 1992, 273 .
\bibitem{W3}
M. Fichtm\"uller, A. Lorek, and J. Wess, \zpc 71, 1996, 533 .
\bibitem{W2}
A. Lorek, A. Ruffing, and J. Wess, \zpc 74, 1997, 369 .
\bibitem{b}
L. C. Biedenharn, \jpa 22, 1989, L873 .
\bibitem{mcf}
A. J. Macfarlane, \jpa 22, 1989, 4581 .
\bibitem{bon}
D. Bonatsos and C. Daskaloyannis, \plb 307, 1993, 100 .
\bibitem{mmp}
S. Meljanac, M. Milekovi\'c, and S. Pallua, \plb 328, 1994, 55 .
\bibitem{w9}
J. Madore, S. Schraml, P. Schupp, and J. Wess,
{\it Gauge Theory on Noncommutative
Spaces}, hep-th/0001203.
\bibitem{W1}
J. Wess, {\it q-deformed Heisenberg Algebras}, math-ph/9910013.
\bibitem {W4}
B. L. Cerchiai, R. Hinterding, J. Madore, and J. Wess, \epjc 8, 1999, 547 .
\bibitem{W5}
B. L. Cerchiai, R. Hinterding, J. Madore, and J. Wess, \epjc 8, 1999, 533 .
\bibitem{W7}
R. Dick, A. Pollok-Narayanan, H. Steinacker, and J. Wess, \epjc 6, 1999, 701 .
\bibitem{BB}
D. S. Bateman, C. Boyd, and B. Dutta-Roy, \ajp 60, 1992, 833 .
\bibitem{bor}
V. V. Borozov, {\it "Orthogonal Polynomials and Generalized Oscillator
Algebras"}, math.CA/0002226.
\bibitem{C}
F. Calogero, \jmp 10, 1969, 2191,2197 ; \jmp 12, 1971, 419 .
\bibitem{pl}
M. Plyushchay, {\it Hidden nonlinear supersymmetries in pure
parabosonic system}, hep-th/9903130.
\bibitem{Brink}
L. Brink, T. H. Hansson, S. Konstein, and M. A. Vasiliev, \npb 401, 1993, 591 .
\bibitem{Poly92}
A. Polychronakos, \prl 69, 1992, 703 .
\bibitem{B2}
R. L. Graham, D. E. Knuth, and O. Patashnik, {\it Concrete Mathematics},
(Adison-Wesley, Reading, Massachusetts, 1989).
\end{document}